\documentclass[twocolumn]{emulateapj}
\usepackage{multirow}
\usepackage{natbib}
\usepackage{xspace}
\usepackage[colorlinks=false,linkcolor=black, citecolor=green, linktoc=page, urlcolor=cyan, 
pdffitwindow=false]{hyperref}
\usepackage[latin1]{inputenc}
\usepackage{amsmath}
\usepackage[usenames,dvipsnames]{color}
\usepackage{rotating}

\def\reff@jnl#1{{\rm#1\/}}
\def\aj{\reff@jnl{AJ}}         
\def\araa{\reff@jnl{ARA\&A}}      
\def\apj{\reff@jnl{ApJ}}        
\def\apjl{\reff@jnl{ApJ}}        
\def\apjs{\reff@jnl{ApJS}}       
\def\aap{\reff@jnl{A\&A}}        
\def\aapr{\reff@jnl{A\&A~Rev.}}     
\def\aaps{\reff@jnl{A\&AS}}       
\def\mnras{\reff@jnl{MNRAS}}      
\def\physrep{\reff@jnl{Physics Reports}}
\def\prd{\reff@jnl{Phys.Rev.D}}     
\def\prl{\reff@jnl{Phys.Rev.Lett}}   
\def\pasp{\reff@jnl{PASP}}       
\def\pasj{\reff@jnl{PASJ}}       
\def\nat{\reff@jnl{Nature}}       
\def\jcap{\reff@jnl{JCAP}}   
\def\memsai{\reff@jnl{MemSAI}} 
\def\na{\reff@jnl{New Astronomy}}       

\def\Sref#1{$\S$\ref{#1}\xspace}

\def\Fref#1{Figure~\ref{#1}\xspace}

\def\Tref#1{Table~\ref{#1}\xspace}

\def\Eref#1{Equation~\ref{#1}\xspace}

\def\Aref#1{Appendix~\ref{#1}\xspace}
\def\Cref#1{Chapter~\ref{#1}\xspace}

\newcommand{\chihway}[1]{\textcolor{black}{#1}}

\begin{document}

\title{Beam calibration of radio telescopes with drones}

\author{Chihway Chang\altaffilmark{*}, Christian Monstein, Alexandre Refregier, Adam Amara \\ 
Adrian Glauser, Sarah Casura \\}
\affil{Institute for Astronomy, Department of Physics, ETH Zurich, Wolfgang-Pauli-Strasse 27, 8093 Z\"urich, Switzerland}
\altaffiltext{*}{Electronic address: chihway.chang@phys.ethz.ch}

\begin{abstract}

We present a multi-frequency far-field beam map for the 5m dish telescope at the Bleien Observatory 
measured using a commercially available drone. We describe the hexacopter drone used in this 
experiment, the design of the flight pattern, and the data analysis scheme. This is the first application of this calibration 
method to a single dish radio telescope in the far-field. The high signal-to-noise data allows us to characterise the beam 
pattern with high accuracy out to at least the 4th side-lobe. The resulting 2D beam pattern is compared with that derived 
from a more traditional calibration approach using an astronomical calibration source. We discuss the advantages of 
this method compared to other beam calibration methods. Our results show that this drone-based technique is very 
promising for ongoing and future radio experiments, where the knowledge of the beam pattern is key to obtaining 
high-accuracy cosmological and astronomical measurements.\\

\end{abstract}


\section{Introduction}
\setcounter{footnote}{0}

In the next decade, a number of large radio experiments are scheduled to begin data collection. One of the key 
science goals of these programmes is to map the HI intensity in the Universe through its 21 cm emission line. 
HI intensity mapping provides a probe of the Baryon Acoustic Oscillation (BAO) feature in the \chihway{matter 
power spectrum (or more directly, the HI power spectrum)} that is independent of traditional measurements 
using galaxy clustering and weak lensing \citep{2008MNRAS.383.1195W, 2010Natur.466..463C, 2015ApJ...803...21B}.
Examples of ongoing efforts in this area include  
the HI Parkes All-Sky Survey \citep[HIPASS,][]{2005MNRAS.359L..30Z},
the HI Jodrell All-Sky Survey \citep[HI-JASS][]{2003MNRAS.342..738L},
and the Arecibo Fast Legacy ALFA Survey \citep[ALFALFA][]{2010ApJ...723.1359M}.
Many more future programmes are also being designed and built, including 
the Square Kilometre Array\footnote{\url{https://www.skatelescope.org/}} (SKA),
the Low Frequency Array\footnote{\url{ http://www.lofar.org}} (LOFAR),
and the Baryon acoustic oscillations In Neutral Gas Observations \citep[BINGO,][]{2012arXiv1209.1041B, 
2013MNRAS.434.1239B}. 

For HI intensity mapping, especially at low redshift, an advantageous survey configuration is to operate dish arrays of small to 
moderate dish sizes (5-15m) in single-dish configurations \citep{2015arXiv150103989S}. This allows wide collecting area 
and a complete sampling of relevant spatial scales. However, in order 
to achieve the required accuracy for the single-dish telescopes, one needs to understand and calibrate the response pattern, 
or the beam of each telescope very well. Small deviations of the mechanical configuration or the environment (e.g. 
temperature, wind) can cause changes to the beam pattern and introduce systematic errors in the measurement. 
\chihway{A 1\% uncertainty in the size of the beam roughly propagates into a 4\% systematic uncertainty in the amplitude 
of the power spectrum. Although current constraints from HI intensity mapping are still dominated by $\sim20\%$ statistical 
errors \citep{2015MNRAS.447.3745P}, with future large surveys where percent-level statistical uncertainties are expected, 
the systematic errors from the knowledge of the telescope beam needs to be controlled to sub-percent level.}

Traditionally, beam calibration has been done using bright astronomical sources such as the sun 
\citep[e.g.][]{1966raas.book.....K}, the moon \citep{2013AA...556A...1T}, and known bright radio sources such as 
Cassiopeia A, Taurus A, Cygnus A, and Virgo A \citep{1977AA....61...99B}. Having the source drift-scan over the 
extent of the beam gives one a measure of the beam shape convolved with the source. 
Similarly, one can use satellites or other artificial sources placed on distant towers to perform 
such calibration. However, in the case of astronomical sources, the number of usable sources is limited and 
decreases for smaller radio dishes. 
Furthermore, the fluxes and the sizes of these sources can fluctuate over time. In the case of 
satellites, the frequency range of the source spectrum is usually very limited, though the intensity is fairly 
high and regular.
To avoid these limitations in using astronomical objects or satellites as calibration sources, the ideal 
solution is to construct a artificial calibration source that is flexible and controllable so one can tailor 
it to the specific telescope and experiment of interest. 

In this paper, we implement this idea by using a noise source carried by a commercial hexacoptor drone. 
The noise source emits a flat spectrum in the frequency range 980 MHz -- 1250 MHz at high power. The 
frequency range is chosen to be the 21 cm frequency redshifted to $z=0.14$-0.46. The drone flies in a 
region where the far-field beam pattern can be mapped. We show that this method gives a controllable, 
light-weight solution to the beam calibration problem for radio telescopes. An earlier 
study\footnote{\url{http://www.skatelescope.org/wp-content/uploads/2011/03/SKA_NEWSLETTER_VOLUME_25.pdf}}
with a similar setup has been done 
by the aperture array verification programme (AAVP) as a proof of principle. 
In this paper we work with a different telescope type and wavelength. We also perform more quantitative 
analyses of the data to show the potential of this method. 
Note that although this work is motivated by HI intensity mapping cosmology, the application of 
our method can be extended to other science areas where single-dish radio telescopes are used in the 
centimeter wavelength, such as solar physics, pulsars and radio bursts.

This paper is organised as the follows. In \Sref{sec:design} we describe the separate components of the 
experiment design. In \Sref{sec:beam_param} we introduce definitions of the characteristic quantities we 
like to measure from our beam. The data processing, analysis and final results of this experiment are 
presented in \Sref{sec:analysis}. We also compare our measurements from the drone with other more 
traditional approaches. In \Sref{sec:conclusion}, we present our conclusions. 

\section{Experiment design}
\label{sec:design}

\begin{figure}
  \begin{center}
  \includegraphics[scale=0.28]{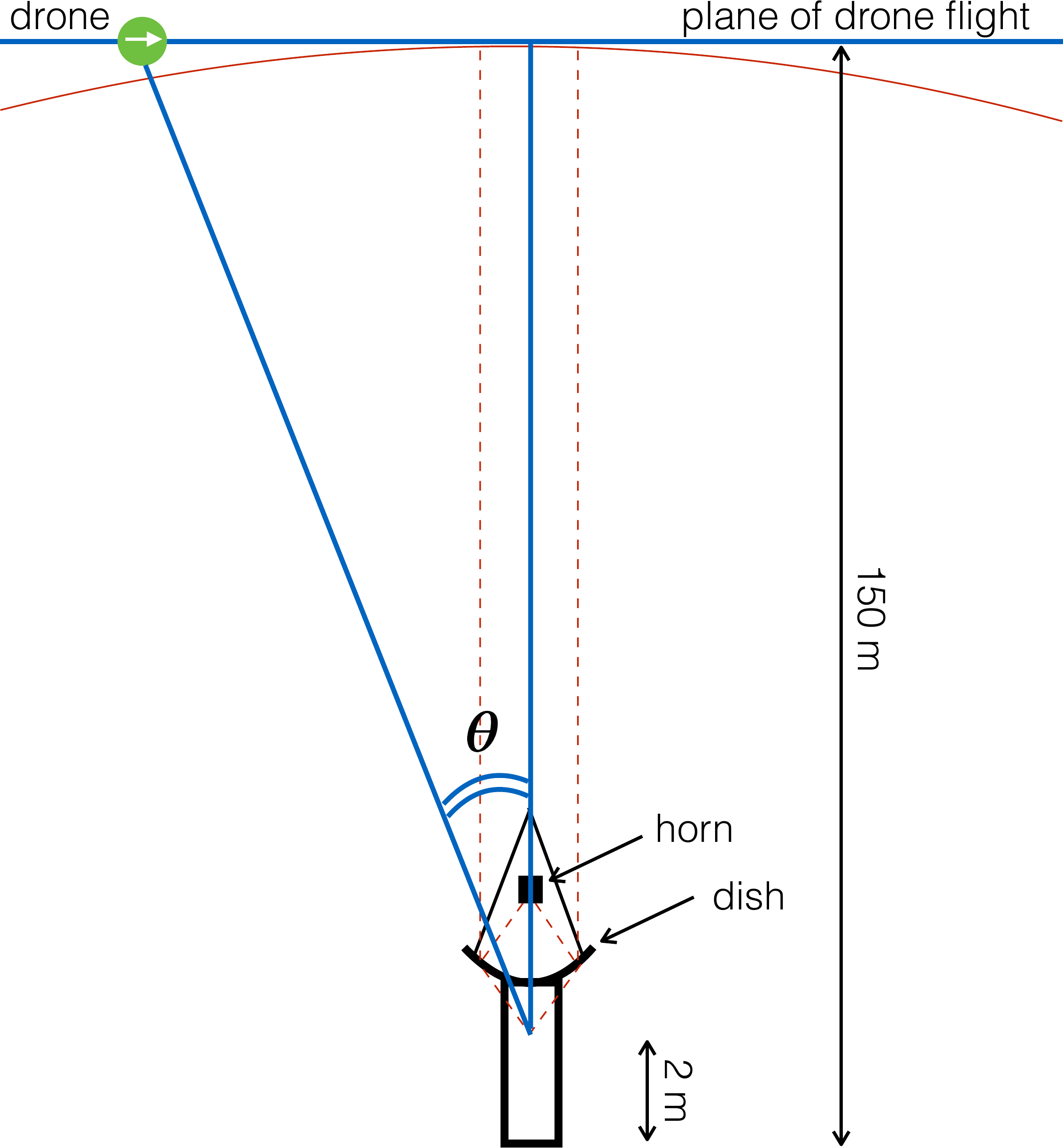}  
  \end{center}
  \caption{Schematics of the experiment setup. The drone, indicated by the green circle, is flying about 150 m above 
  ground, in a plane directly above the telescope. 
  Note that the dimensions in the plot are not drawn to-scale.}
\label{fig:exp_setup}
\end{figure}

We describe below the separate components of the experiment: the telescope, the spectrometer, the 
drone, the noise transmitter, and the design of the flight pattern. \Fref{fig:exp_setup} illustrates the 
schematics of the experiment setup. A test run of smaller scale was performed on October 28, 2014, 
while the full experiment was carried out on November 20, 2014. All data presented in this paper are 
from the latter dataset.  

\vspace{0.2in}

\subsection{The Bleien 5m dish}
\label{sec:telescope}

\begin{figure}
  \begin{center}
  \includegraphics[scale=1.0]{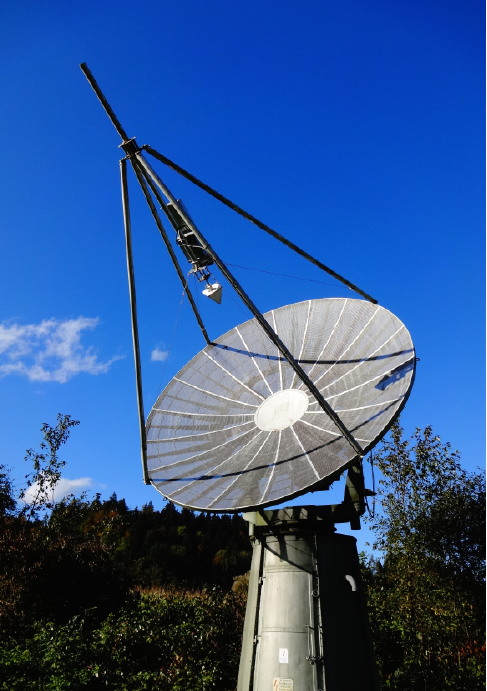}  
  \end{center}
  \caption{Image of the 5m radio dish at the Bleien Observatory. From the image one can see the wire meshed 
  reflector, as well as the four struts which hold the horn receiver in focus.}
\label{fig:dish_photo}
\end{figure}
\vspace{0.2in}

We demonstrate our new calibration technique on the 5m parabolic dish\footnote{The radio dish was constructed 
in 1972 by the Swiss company Schweizerische Wagons und Aufzügefabrik AG.} (f/D$=0.507$) at the Bleien 
Observatory\footnote{\url{http://www.astro.ethz.ch/research/Facilities/Radioteleskop_Bleien}} in Gränichen, 
Switzerland (longitude 8.112215$^{\circ}$, latitude 47.3412278, altitude 469 m). The surrounding area 
$\sim 1.5$ km in radius is protected against commercial radio emission in the frequency range of interest for 
this experiment. \Fref{fig:dish_photo} shows an image of the telescope. During the experiment, the telescope 
is pointed vertically up towards the zenith so that the beam pattern can be measured at a plane parallel to the 
ground. As discussed in \Sref{sec:procon}, the telescope can also be pointed at different elevation angles to 
map the beam shape as a function of elevation. A cylindrical horn feed is supported via four struts at the focal 
point. The cylindrical horn has a length of 185 mm, diameter of 200 mm and the dipole length is 85 mm. 
 
\begin{figure*}
  \begin{center}
  \includegraphics[scale=1.0]{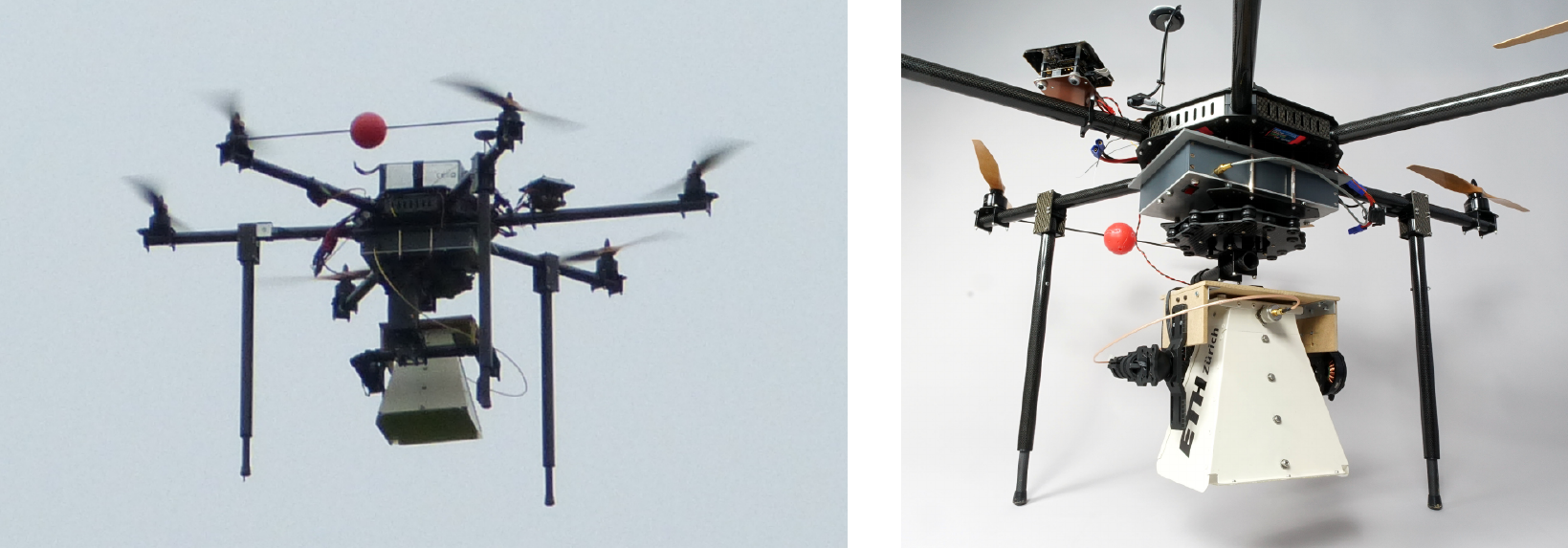}  
  \end{center}
  \caption{Image of the drone and noise transmitter horn used in this experiment. The left image is taken during 
  the flight and the right image shows a zoom-in of the vehicle, transmitter and gimbal. The red ball on the drone 
  is the ``nose'', which helps the user to identify the orientation of the vehicle. The noise transmitter horn can be 
  seen held by the gimbal and pointing vertically down at all times.}
\label{fig:drone_photo}
\end{figure*}

\subsection{The CALLISTO spectrometer}

During the experiment, data is collected by the CALLISTO 
spectrometer \citep{Benz:2005aa}. The CALLISTO spectrometer is a programmable heterodyne 
receiver built in the framework of IHY2007 and ISWI by former Radio and Plasma Physics Group at 
ETH Zurich, Switzerland. The instrument natively operates between 45 and 870 MHz and has a 
frequency step size of 62.5 kHz. The data from the telescope is down-converted to match the frequency 
range of CALLISTO.  The data obtained from CALLISTO are FITS-files with up to 400 
frequencies per sweep. We set the time resolution of the data to be 0.25 sec and 200 channels per 
spectrum. The integration time is 1 ms, \chihway{the radiometric bandwidth is $\sim$300 kHz ($\sim5$ times the 
frequency step size)}, and the overall dynamic range is larger than 50 dB. 
 
\subsection{Drone}

\begin{deluxetable}{ll}
\tablewidth{0pt}
\centering
\tablecaption{Basic characteristics of the drone.} 
\tablehead{Quantity & Specification} 
\startdata
Diameter of full vehicle  &   110 cm        \\
Weight                & 10.88 kg$^{a}$ (total)        \\
Maximum motor power     & 2.01 kW         \\
Propeller dimensions & 16" (diameter) $\times$ 6" (pitch) \\
Flight control system &    DJI WooKong-M$^{b}$ \\
Maximum flight duration & 13.5 minutes        
\enddata
\tablenotetext{a}{Including 2.73 kg for the weight of the accumulator.}
\tablenotetext{b}{\url{http://www.dji.com/product/wookong-m}}
\label{tab:drone_params}
\end{deluxetable}
\vspace{0.2in}

The hexacopter drone used in this experiment was provided by the private company 
Koptershop\footnote{\url{http://www.koptershop.ch/}}. The main characteristics of the vehicle are listed in 
\Tref{tab:drone_params}, while an image of it is shown in \Fref{fig:drone_photo}. The critical features 
of the drone considered in this work are the following:
\begin{itemize}
\item The drone should be able to carry the weight of the noise transmitter in addition to its own weight.
\item The gimbal on the drone should be able to steadily point the noise transmitter to a given direction, 
which means that the drone flight needs to be stable, and the gimbal should compensate for any instability.
\item The drone should be able to sustain a flight long enough for at least one pass through the expected 
beam pattern. 
\item The 3D position of the drone should be recorded to sufficient accuracy when the signal is transmitted 
from the noise transmitter. 
\end{itemize}
All the above requirements can be met with commercially available drone vehicles. Specifically, the gimbal used 
was purchased commercially and modified to fit the specifications of the experiment. For the last point above, the 
position of the drone during the flight is given by 5 or more GPS satellites (depending on the situation 
during the flight). The GPS absolute positioning is accurate to a few meters in the transverse direction (we 
estimate the relative accuracy in \Sref{sec:analysis}). The height of the drone position is controlled by a 
barometric altimeter carried by the drone and is re-calibrated before each flight. The barometric altimeter 
measurements are typically accurate to well within a meter.   

\subsection{Noise transmitter}

The noise transmitter is composed of a non-coherent semiconductor noise source, an attenuator, a 
broad-band amplifier, a band-pass filter, a power supply and a transmission antenna. The whole unit is 
light-weight ($<2$ kg), making it possible to be carried by the drone for extended flights. 

The band-pass filter ensures that only the 
frequency range of interest is transmitted. 
Approximately 3W of total power is needed 
for the noise transmitter, which is separate from the power supply for the drone. Finally, the transmission 
antenna is a double ridged horn antenna as shown in the right panel of \Fref{fig:drone_photo}. The antenna 
is constructed with thin light-weight aluminium sheets 
and covered by a polystyrene plane, with gain of maximum 5 dB and a frequency range similar to our 
band of interest.
The antenna is linearly polarised, with a fluxgate magnetometer to maintain the stability of the polarisation.

Note that to transmit in our frequency band at such intensity in Switzerland, a transmission permit from the 
Federal Office of Communications (OFCOM) was obtained.

\subsection{Design of the flight pattern}

The flight pattern is designed to fully cover the extent of the beam at far-field, while having sufficient resolution of the 
high-order side lobes. We fly at an elevation of about 150 m. At the frequency of interest ($\sim 1$ GHz), this elevation 
is sufficiently close to the far-field region defined by the Fraunhofer distance:
\begin{equation}
d_{f} = \frac{2D^{2}}{\lambda} \approx 166\: {\rm m},
\label{eq:farfield}
\end{equation}
where $D$ is the telescope aperture and $\lambda$ is the wavelength of interest. \Eref{eq:farfield} suggests 
that for larger telescopes and shorter wavelengths, the far-field requirement is more stringent. The commercial drone 
we used provide an appropriate platform for testing the Bleien 5m telescope at these wavelengths. For experiments with 
a much larger far-field requirement, more advanced drones can be used. Alternatively, one can carry out the experiment 
in near field and reconstruct the far-field beam via modelling (an example modelling software is introduced in 
\Aref{sec:grasp}).

Given the elevation mentioned above, we set the flight pattern to be on a rectangular grid of 75 m$\times$75 m 
directly above the telescope, which comfortably covers the beam out to the fourth side lobe. The grid is oriented in the 
North-South (NS)/East-West (EW) direction with each flight track separated by 5 m. The 5 m spacing corresponds to 
about 1.9$^{\circ}$ at the beam centre ($\sim 0.5$ times the FWHM of the beam at 1GHz), suggesting roughly three 
tracks would pass through the extent of the main beam. This means there are 16 flight tracks in 
the NS direction and 16 in the EW direction. The flight time limit of the drone allows it to complete two tracks for 
each flight, then the batteries need to be changed. The polarisation angle of the reciever feed horn 
is 45$^{\circ}$ in the NW-SE direction, while the polarisation angle of the noise transmitter is always parallel to the 
direction of flight. This suggests that there should be no amplitude differences for the tracks along the NS and EW 
tracks due to the polarisation. The left panel of \Fref{fig:flight_path} illustrates the schematics of the raster scan pattern 
for this experiment. The full flight pattern is programmed into the drone control software so that it runs automatically. 
Manual control is invoked only during takeoff and landing for safety considerations. 

\section{Beam characterisation}
\label{sec:beam_param}

The radio beam prescribes the sensitivity of a radio telescope as a function of the angle of the incoming ray relative to 
the telescope pointing. Typical beam profiles are composed of a prominent main beam and side lobes. The effect of the 
side-lobes is to pick up signals that are not in the direction of interest. The larger the main beam is relative 
to the side lobes the more efficient the beam is in collecting signal. The narrower the spatial extent of the main beam, 
the higher its resolution. The goal of this paper is to map the 3D (2D in angular space + frequency) beam of the Bleien 5m 
radio dish. From the beam map, we also derive basic characteristics of the telescope. 

We follow the terminology used in \citet{1986tra..book.....R}.
Assume $P(\vec{\theta}; \lambda)$ to be the normalised beam with peak intensity equal to one and falls to zero at 
infinity. The first convenient measurement is the Full-Width-Half-Maximum (FWHM) of the main beam, or the average 
diameter of the contour where $P(\vec{\theta}; \lambda)=0.5$. In an ideal case with only a perfect circular top-hat 
aperture, the FWHM depends on the wavelength $\lambda$ and the dish diameter $D$ according to
\begin{equation}
{\rm FWHM(\lambda)} = 1.028 \frac{\lambda}{D}.
\label{eq:fwhm}
\end{equation}

Second, we can integrate $P$ over the full 4$\pi$ solid angle to get the beam solid angle $\Omega_{A}$, or integrate 
only inside the first null to get the main beam solid angle $\Omega_{M}$.
\begin{equation}
\Omega_{A}(\lambda) = \int\int_{4\pi} P(\vec{\theta};\lambda) d^{2} \theta;
\label{eq:Oa}
\end{equation}
\begin{equation}
\Omega_{M}(\lambda) = \int\int_{\rm main \; lobe} P(\vec{\theta};\lambda) d^{2} \theta.
\label{eq:Om}
\end{equation}

From $\Omega_{A}$ and $\Omega_{M}$ we can calculate two other quantities. 
The beam efficiency is defined as
\begin{equation}
\eta_{M}(\lambda) = \frac{\Omega_{M}(\lambda)}{\Omega_{A}(\lambda)}.
\label{eq:Em}
\end{equation}
$\eta_{M}(\lambda)$ is a measure of the relative level between the main beam and the side lobes. The closer 
$\eta_{M}(\lambda) $ is to 1, the more prominent the main beam is 
and the more efficient the beam is in collecting the signal.  
The effective aperture of a beam is defined as 
\begin{equation}
A_{e}(\lambda) = \frac{\lambda^{2}}{\Omega_{A}}.
\end{equation}
The aperture efficiency is defined as the ratio of $A_{e}$ to the geometric aperture $A_{g}=\pi (D/2)^2$, or 
\begin{equation}
\eta_{A}(\lambda) = \frac{A_{e}}{A_{g}} = \frac{4 \lambda^{2}}{\pi \Omega_{A} D^{2}}.
\label{eq:Ea}
\end{equation}

In \Sref{sec:wav_dep}, we calculate ${\rm FWHM(\lambda)} $, $\eta_{M}(\lambda)$ and $\eta_{A}(\lambda)$ 
for our beam measurement.
 
\section{Analysis and results}
\label{sec:analysis}

In this section we describe the analysis procedure and show the results of our beam measurement. We first 
describe the data processing steps in \Sref{sec:data_process}. Next we present the results in terms of the 1D profile 
(\Sref{sec:1d}), 2D profile (\Sref{sec:2d}), and wavelength-dependence (\Sref{sec:wav_dep}) of the beam. Finally we 
compare the drone measurements with other approaches in \Sref{sec:compare}. 

\subsection{Data processing}
\label{sec:data_process}

\begin{figure*}
  \begin{center}
  \includegraphics[scale=0.32]{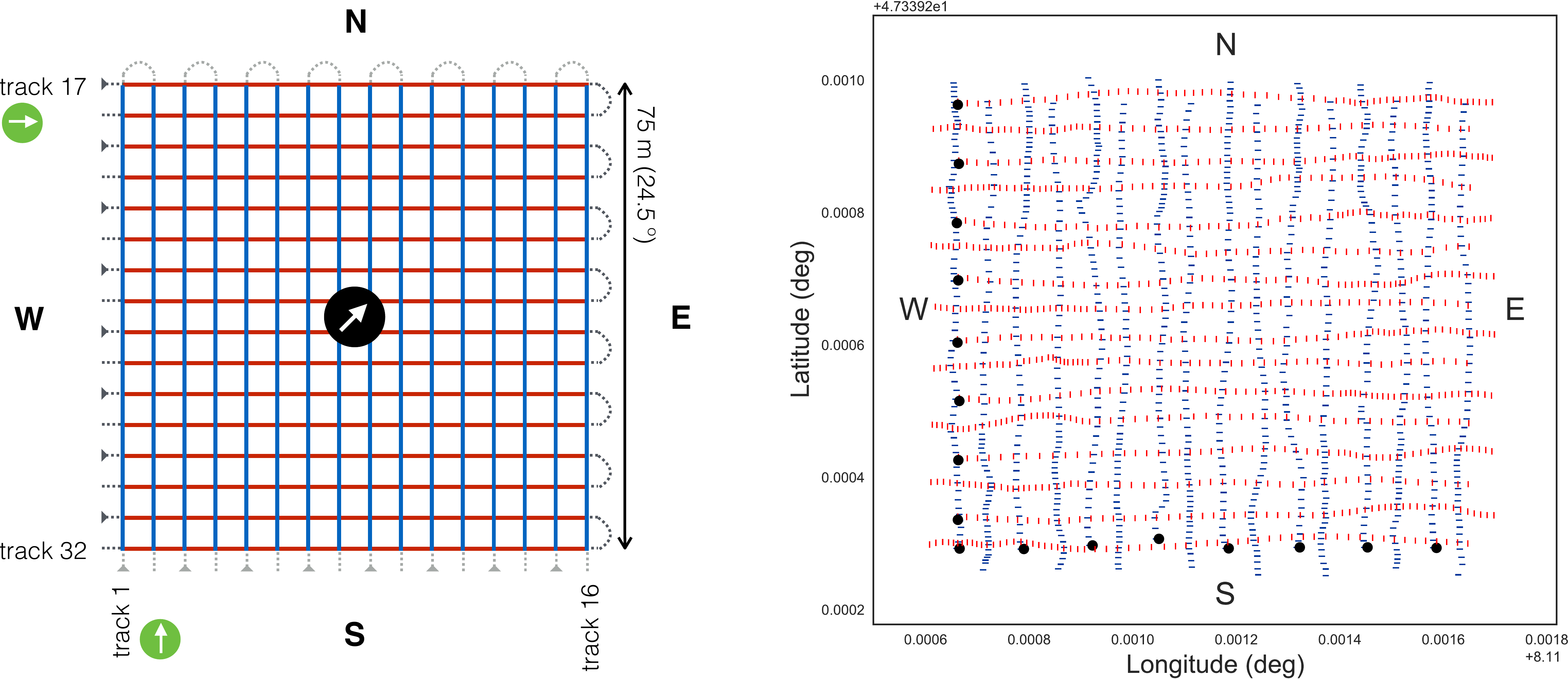}  
  \end{center}
  \caption{Schematics of the flight pattern design (left) and the actual flight path during the experiment recorded by 
  GPS (right). In the left panel, the black circle in the middle indicates the telescope, with the arrow showing the polarisation 
  direction of the feed horn on the telescope. The blue (red) lines show the flight path in the NS (EW) orientation 
  where signal from the noise transmitter is emitted. The dashed lines show the flight path where signal is not emitted. 
  The triangle at the beginning of each dashed line shows the beginning of each flight, which include two tracks. The 
  green circles and the arrows within indicates the drone and the orientation of the noise transmitter polarisation in 
  the NS (lower green circle) and EW (upper green circle) direction. In the right panel, each stroke indicates an 
  emission from the noise transmitter. The blue strokes are in the NS direction while the red strokes are the in EW 
  direction.The black dots indicate the beginning of each track, corresponding to the triangles in the left panel. One can 
  see that the distance between each stroke is not constant due to the change in the drone's velocity. }
\label{fig:flight_path}
\end{figure*}

From the spectrometer, we read out time-series signal from the receiver over 200 frequency channels. The first 
task is to match the signal from the spectrometer at every time-instant to the drone location in the air when this 
signal was emitted. From the GPS data, we have a coordinate record for each emitted signal, which includes 
2-sec ``on'' and 1-sec ``off'' signal from the noise transmitter. The right panel of \Fref{fig:flight_path} shows the actual 
GPS records for the position of the drone at each pulse from the noise transmitter. We calculate the median 
RMS scatter in the longitude (latitude) direction for all NS (EW) tracks to be 0.55 (0.51) meters, which corresponds 
to 0.21$^{\circ}$ (0.2$^{\circ}$) at the beam centre, or $\sim5\%$ of the beam. All the emitted signals are recorded by 
the spectrometer, suggesting 
none of the emission signals were too weak to be detected. As the drone does not fly at a constant speed, the distance 
between each on-off signal changes. We take this into account when assigning a coordinate in the air to each 
of the spectrometer data pixels. 

\begin{figure*}
  \begin{center}
  \includegraphics[scale=0.45]{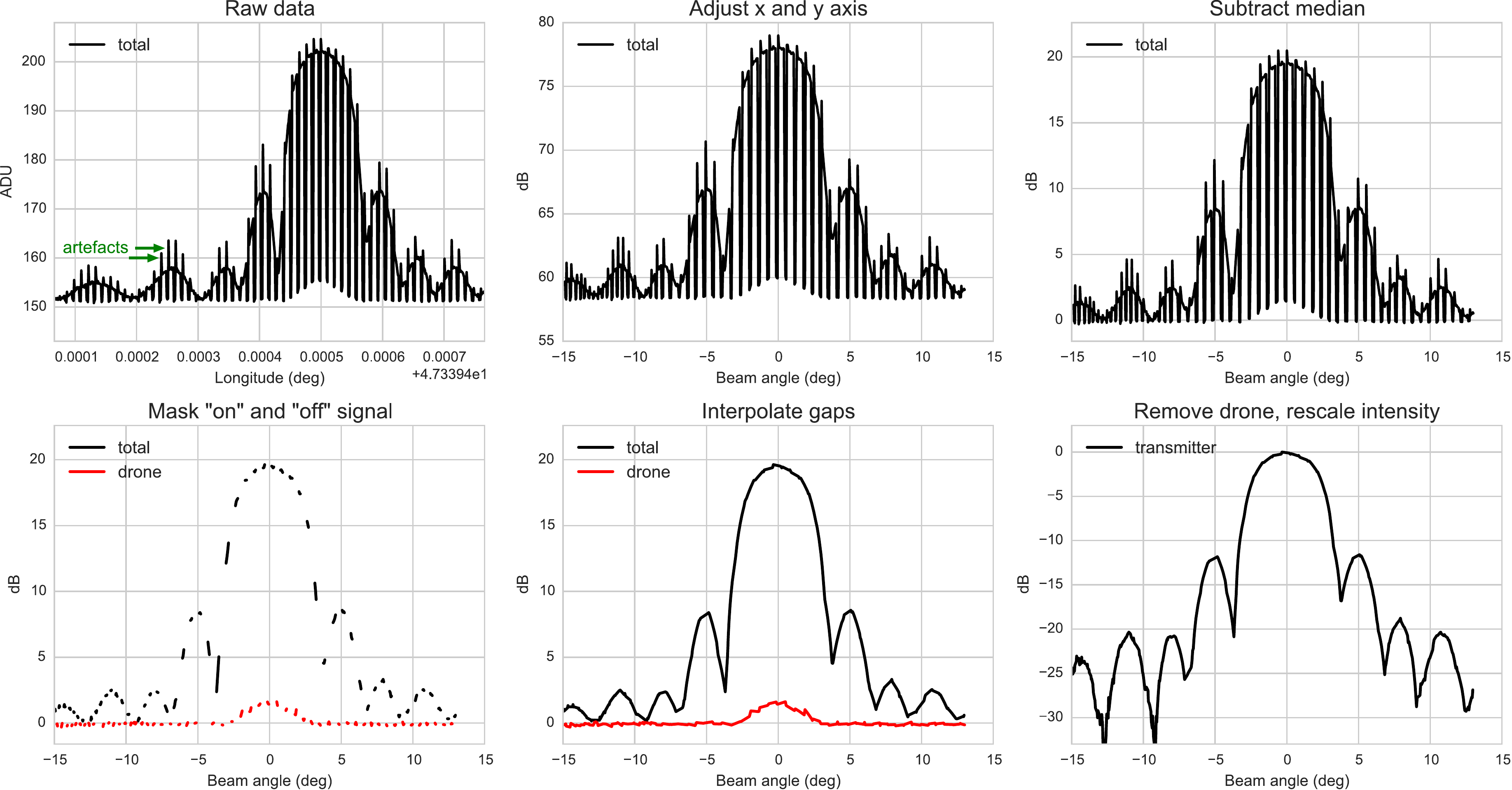}  
  \end{center}
  \caption{Example of the data processing for the NS track closest to the beam centre. These curves correspond to a  
  stacking of the frequencies in the range 1191.9 -- 1220.2 MHz (20 channels with mean frequency 1206 MHz). Top 
  row from left to right shows the raw data, the data adjusted for the units in both axes, the data corrected for the 
  standing-wave pattern. The green arrows on the top left panel indicates the artefacts induced by the transmitter signal. 
  Bottom row from left to right shows the data masking out the 'on' and 'off' signal separately, 
  interpolation over the data gaps, and the final corrected beam profile.  }
\label{fig:data_process}
\end{figure*}

The raw data from the spectrometer appears as a series of on-off signals, with often artefacts at the edge of 
the off transition due to the electronics in the transmitter as shown by the green arrows in the top left panel 
of \Fref{fig:data_process}. \chihway{The on-off signal is designed to help remove the background and low-level RFI from 
the signal of interest as explained below.} The data is cleaned via the following steps:
\begin{enumerate}
\item Convert the units of the raw data into dB\footnote{Note that, in this experiment, we can only 
measure the relative level of the beam intensity at each position, as the instrument noise and throughput is 
not calibrated. In the future with better characterisation of the noise transmitter, we can consider doing an 
absolute (radiometric) measurement.} by multiplying the data with the conversion factor used in the CALLISTO 
spectrometer
$\frac{2500 {\rm (mV)}}{{255 {\rm (ADU)} 25.4 {\rm (mV/dB)}}}$.
\item Match the GPS positions (in longitude and latitude) to the signal received by the spectrometer, then 
convert the longitude and latitude to angles from the optical axis of the beam. 
\item For each frequency, subtract all data by the median value over time. This step removes the 
time-independent low-level standing-wave pattern 
(see discussion in \Sref{sec:wav_dep}). 
\item Remove the ``off'' signal and the artefacts around the ``off'' signal by first placing a cut at 0.1 dB and then 
removing 4 pixels on each side around the cut. Manually mask any remaining ``off'' signals that were not cut 
out\footnote{These are typically regions where the interference from the drone actually raises the ``off'' signal 
to a non-negligible level.}.   
\item Linearly interpolate over the gaps in the signal.
\item Using a similar approach as above, we can remove the ``on'' signal to get the background, including the 
interference from the drone motors.
\item Subtract the interference from the drone motor from the total signal, and rescale the amplitude so that the 
peak is at 0 dB.
\end{enumerate}
\Fref{fig:data_process} illustrates these different steps.  

Two geometrical issues also need to be considered to calibrate the reconstructed beam. First, we 
check that the height of the drone during the flight has been stable within 10 cm (upper bound of barometric 
altimeter precision), this corresponds to a $<$0.3\% change in intensity. Second, as the beam pattern is measured on a 
tangent plane at the centre of the beam, one needs to account for the free-space lost between the tangent plane 
and the sphere centred at the telescope and touching the plane (as shown in \Fref{fig:exp_setup}). Both effects 
are negligibly small with the current experimental configuration. 

After calibrating all 32 tracks as described above, we have now a 2D plane with information about the beam on the 
grid formed by the tracks. We interpolate this plane using the \texttt{python} function
\texttt{scipy.interpolate.Rbf} and \texttt{epsilon=1} to form a 2D map of the beam. The map is done for different 
frequency bins.

\subsection{1D beam pattern}
\label{sec:1d}

\begin{figure*}
  \begin{center}
  \includegraphics[scale=0.35]{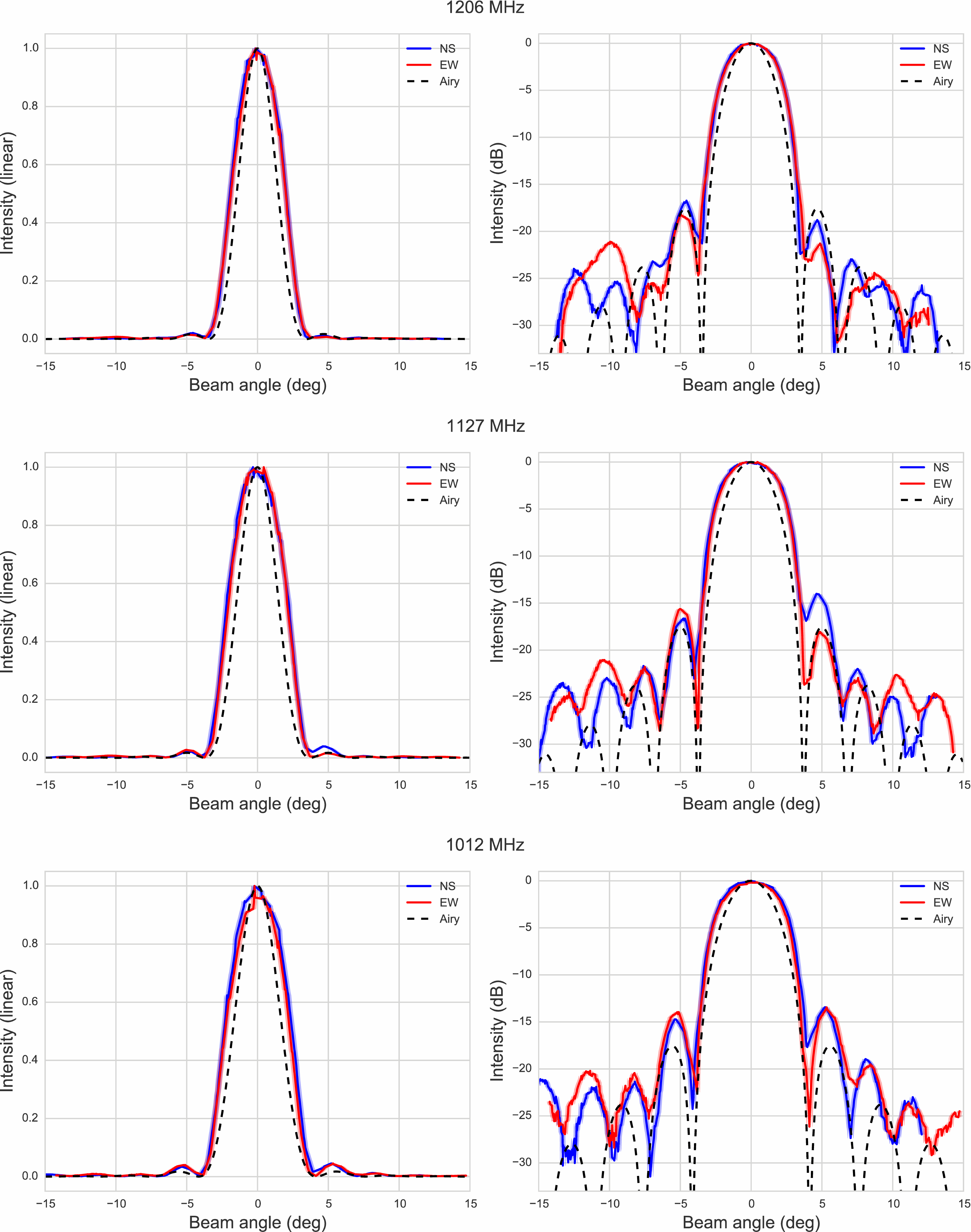}  
  \end{center}
  \caption{The centre beam profile measured in the NS and EW directions in linear (left) and log (right) scales for different 
  frequencies. Each measurement curve is an average of 20 frequency channels, with the mean frequency being 1206 MHz (top), 
  1127 MHz (middle) and 1012 MHz (bottom) respectively. Airy patterns corresponding to an idealised 5m aperture at the 
  mean frequencies are shown by the black dashed curves. The width of the blue and red line indicate expected uncertainties from 
  the imperfect GPS positioning. }
\label{fig:1d}
\end{figure*}

The 1D beam pattern of the two tracks with maximum intensity is shown in \Fref{fig:1d}. Each panel is constructed 
from an average of 20 frequency channels, with the mean frequency listed in the figure (1206 MHz, 1127 MHz and 
1012 MHz, respectively). To guide the eye, the Airy pattern \citep{1838AnP...121...86A} expected for an idealised 
5m aperture with uniform illumination is overlaid in each panel with the black dashed curve. 
The thickness of the measurement curve corresponds to a 0.5 m error in the GPS positioning.

First, it is worth noting that the measurements are at very high signal-to-noise even at the edge of the measurement 
(4th side-lobe). In principle, one can measure the beam pattern to further out with the current setup. 
We find that the Airy pattern gives a main beam size smaller than the drone measurement. This is expected as any 
de-focusing and aberration problems caused by imperfect geometry of the feed horn and the dish tend to enlarge the 
main beam. The nulls are in general not as deep as that predicted from the Airy pattern, but the positions of the first 
null and first side lobe agrees quite well. The positions of the higher-order side lobes and nulls in the measurement 
are shifted towards the main beam compared to the Airy pattern. 
The two measurements from the NS and EW tracks agree fairly well in the position and level of the peak/nulls, while 
the measurements show that the beam is not entirely symmetric. The asymmetry can be due to the intrinsic 
asymmetry of the telescope structure -- the front-end unit is mounted slightly off-axis, and the dipole inside the feed 
horn is not centred. Finally, the beam size increases going to longer wavelengths, as expected. We discuss further in 
\Sref{sec:wav_dep} these wavelength-dependent characteristics.

\chihway{In \Aref{sec:grasp}, we invoke a simple antenna modelling tool \texttt{GRASP} to investigate the effect on 
the beam shape when the beam of the feed horn is included, using the geometry of the receiver horn described in 
\Sref{sec:telescope}. We also look at the impact on the beam shape from changes of other model parameters of the 
telescope.}

\subsection{2D beam pattern}
\label{sec:2d}

\begin{figure}
  \begin{center}
  \includegraphics[scale=0.8]{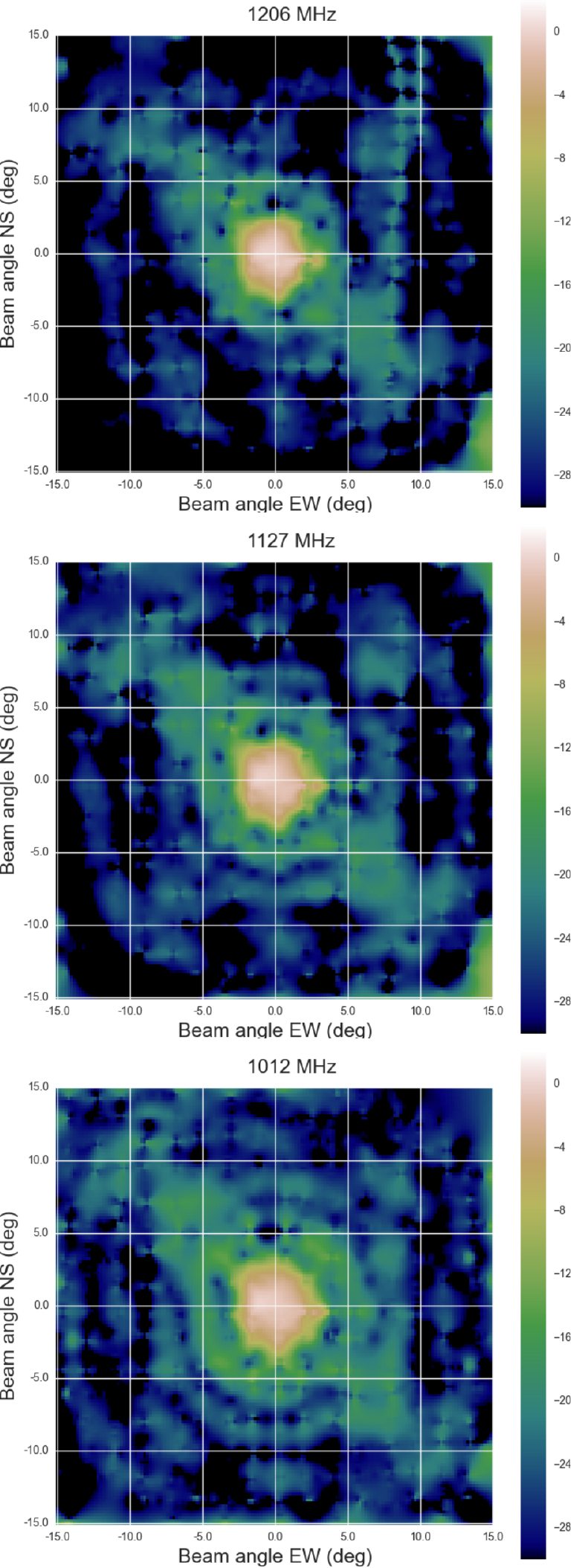}  
  \end{center}
  \caption{Reconstructed 2D beam intensity pattern from drone measurements for different frequencies. Each map is 
  an average of 20 frequency channels, with the mean frequency being 1206 MHz (top), 1127 MHz (middle) and 
  1012 MHz (bottom) respectively. The figure has the same orientation as \Fref{fig:flight_path}. Colour scales are  
  logarithmic (dB).}
\label{fig:2d_map}
\end{figure}

The reconstructed 2D beam pattern is shown in \Fref{fig:2d_map} for the same frequency ranges as \Fref{fig:1d}.  
Visually, one can see that the main beam is centred and roughly circular. The side-lobes show up as concentric ring structures 
centred on the peak of the main beam, with dark rings indicating the nulls. There is noticeable asymmetry along the 45$^{\circ}$ 
direction. This is likely due to the polarisation angle of the telescope (along the 45$^{\circ}$ NE-SW direction 
as shown in \Fref{fig:flight_path}), breaking the otherwise isotropic beam pattern. \chihway{The 4 supporting struts could also introduce 
some of the structures in the beam pattern, as tehy are either parallel or perpendicular to this direction.} Given the flight pattern 
design we used in this experiment (shown in \Fref{fig:flight_path}), artefacts from the grid-pattern are inevitable. That is, we only 
have data taken in certain stripes in the 2D plane, causing the reconstruction to be limited in between the stripes, even 
if there were fine structures in the beam pattern. A more sophisticated flight pattern with adaptive grid size adjusted 
to the expected beam structure can potentially solve this issue in the future. 

These 2D beam maps can be generated for arbitrary frequency ranges and used as input to realistic simulations 
of the sky observed by the telescope.
 
\subsection{Other uncertainties}

We discuss and quantify here the possible sources of uncertainties in our results that we have not considered above. 
All of these effects are subdominant to that coming from the GPS uncertainty in the transverse direction.

\begin{itemize}
\item{\textbf{Gimbal position/angle uncertainty:}}
We measured the stability of the angle of the gimbal to be within 1 deg, this corresponds to a $<$1 deg uncertainty 
in the polarisation of the emission from the transmitter, resulting in $<$0.1\% uncertainty in the flux. The flux attenuation 
from the beam not directly pointing at the dish corresponds to a $<$0.05\% uncertainty in the flux level. 
\item{\textbf{Radio Frequency Interference (RFI) removal:}}
No severe high-level RFI contamination was observed during the flight, while low-level, long time-scale RFI is 
removed during the final step described in \Sref{sec:data_process}.
\item{\textbf{Polarisation uncertainty:}}
One possibile error would arise if the flight path of the drone was not precisely 45 deg from the polarisation angle 
of the telescope. This would appear as a difference in the measured peak of the beam measured from the two main NS 
and EW tracks. We find this difference to be $<$4\%. Note, however, that this number also contains the uncertainties 
in the background subtraction, and other effects mentioned above.
\item{\textbf{Beam shape of transmitter horn:}}
If the beam size of the transmitter horn is too small, it can imprint onto the beam measured from the telescope. 
We estimate this effect using a simple horn antenna model \citep{1950ante.book.....K}. We find that for the tracks furthest 
away from the dish centre, the flux attenuation is about 8\%, 
for the centre region where the main beam is probed (10 meters from the centre of the dish), the attenuation is below 1\%. 
This means that we are underestimating the high-order slide lobes slightly, but the effect of the overall beam size 
measurement is small. 
\end{itemize}

\subsection{Wavelength dependence}
\label{sec:wav_dep}

In this section we calculate the wavelength-dependent beam characteristics from the measured beam profile. We 
take the two centre tracks used in \Fref{fig:1d} and keep all the frequency channels separate. Twenty lowest 
frequency channels were discarded due to severe RFI. The remaining frequency 
range is 997 -- 1256 MHz.

The beam FWHM (\Eref{eq:fwhm}) is estimated by the FWHM of the best-fit Gaussian of the 1D linear normalised 
profile. $\Omega_{A}$ (\Eref{eq:Oa}) and $\Omega_{M}$ (\Eref{eq:Om}) are calculated by integrating under the 
normalised beam map.
In reality, we can only integrate \Eref{eq:Oa} inside our map (the inner 15 deg area of the beam). 
To account for the un-measured regions outside our map, we estimate the fraction of the beam inside our map 
from the beam model described in \Aref{sec:grasp}. We find that $\sim85\%$ of the beam is inside the 15 deg area. 
We thus multiply the measured integration  of our maps 1/0.85$=$1.18 to yield the $\Omega_{A}$ used in \Fref{fig:1d}.
As the first null in the single-frequency maps are often too noisy to define, for $\Omega_{M}$ 
we use the 1D FWHM measurement to estimate the location of the first null in the case of an Airy pattern. From 
\Fref{fig:1d} we can see that the position of the first null is quite well predicted by this approximation.
Finally, $\eta_{A}$ and $\eta_{M}$ can be calculated according to \Eref{eq:Ea} and \Eref{eq:Em}.

\begin{figure}
  \begin{center}
  \includegraphics[scale=0.5]{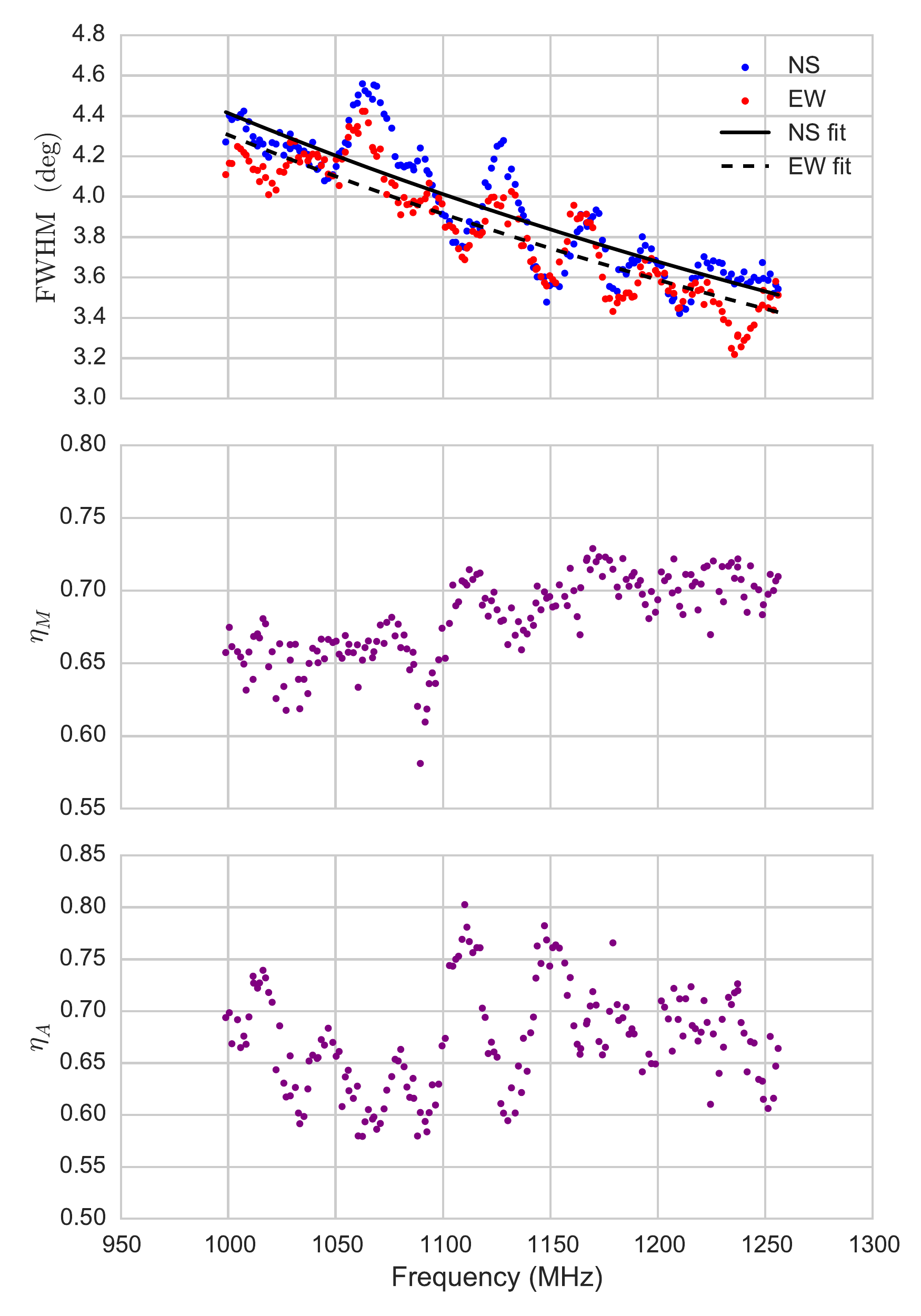}  
  \end{center}
  \caption{Frequency-dependence of the beam characteristics. From top to bottom we plot the FWHM, beam efficiency and 
  aperture efficiency as a function of frequency. The FWHM measurement is done on the 1D profiles of the NS (blue) and 
  EW (red) main track, while $\eta_{M}$ and $\eta_{A}$ are calculated from the full 2D maps.}
\label{fig:wave_dep}
\end{figure}

The measured $\rm FWHM$ in the NS and EW directions as a function of frequency is shown in the first panel of 
\Fref{fig:wave_dep}. Also overlaid are the linear fits to both sets of data points. The fitted slopes are 1.28 and 1.25 
for the NS and EW tracks respectively, which is about $20\%$ larger than the idealised case (\Eref{eq:fwhm}).
The plot also shows a prominent modulation along the linear relation. This modulation matches the time-independent 
standing-wave pattern observed in the data even without any signal present. These standing waves could be introduced 
by multiple reflections off certain structures in the system \citep{1997PASA...14...37B, 2008AA...479..903P}. 
In our case, this is a combination of reflections from physical components (e.g. dish-horn) and reflections within the 
electronics. As shown in \citet{2008AA...479..903P}, the standing-wave imprints through the beam-size measurement, 
which is what we observe in the data. 

The aperture efficiency $\eta_{M}$ and the beam efficiency $\eta_{A}$ as a function of frequency are shown in the 
second and third panel of \Fref{fig:wave_dep}. The median $\eta_{M}$ over this frequency range is $68\%$ 
and the median $\eta_{A}$ is $67\%$. $\eta_{A}$ is fairly constant over frequency, while $\eta_{M}$ increases 
by about 10\% in our frequency range. Both plots also show imprints of the standing wave pattern, which 
appears to be out-of-phase with the FWHM measurement. This is expected, as a larger FWHM would lead to 
a smaller aperture/beam efficiency.

\vspace{0.2in}

\subsection{Comparison with other measurements}
\label{sec:compare}

In this section we compare the drone measurements with a more traditional approach carried out at a time closely following 
the drone measurement, assuming that the beam shape is stable over that time period. It is also important that the instrument 
settings were kept the same as that used in the previous case. In \Sref{sec:sun} we repeat the measurement using the sun 
as a calibration source. We then discuss the pros and cons of the different approaches in \Sref{sec:procon}.    

\subsubsection{Beam measurement with the sun}
\label{sec:sun}

We performed the following sun scan on December 19, 2014. The scanning strategy is designed to be similar to that 
of the drone measurement and allows us to reconstruct the beam pattern in both NS and EW directions. 

\begin{figure*}
  \begin{center}
  \includegraphics[scale=0.5]{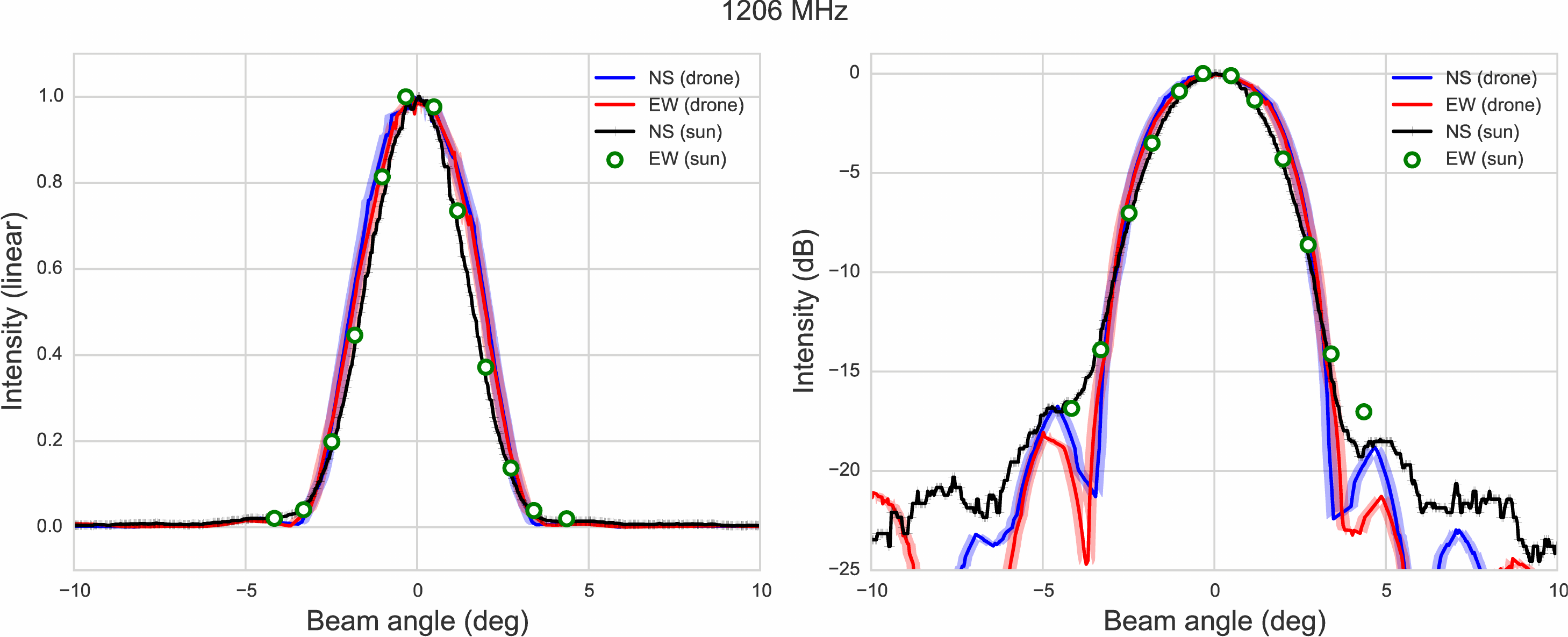}
  \end{center}
  \caption{Linear (left) and log (right) beam profiles in the NS and EW cross sections through the beam centre as 
  measured by the drone and the sun scan. The measurements correspond to an average of 20 frequency channels 
  with the mean frequency 1206 MHz (same as the top panel in \Fref{fig:1d}). For the sun scan, the NS direction has 
  a much lower sample rate due to the scanning strategy.}
\label{fig:sun_drone}
\end{figure*}

The data is taken with the telescope pointing South (azimuth 180$^{\circ}$) and constantly changing elevation up and 
down from $\sim$7$^{\circ}$ to $\sim$32$^{\circ}$  elevation, the sun passes at elevation around 20$^{\circ}$ while going 
east to west through azimuth 180$^{\circ}$. There are about 12 encounters of the sun and the beam where we see visible 
peak in the data. Zooming in each peak, one can see a full smooth profile from scanning the beam in the vertical direction, 
where the angular difference between the pixels is determined by the speed of the telescope slew. The peak of the profiles 
can be identified as the point where the sun passed through the centre of that beam profile. Connecting all the peaks thus 
gives us the centre beam profile in the EW direction. \Fref{fig:sun_drone} shows the sun scan results of the two cross 
sections through the beam centre at a given frequency compared to the drone measurements. Results from other 
frequencies are similar.

\chihway{From \Fref{fig:sun_drone}, we find that both the NS and the EW-beam from the sun are broadly consistent with the drone 
measurements at the 1-2$\sigma$ level. The SNR of the sun measurement is too low to resolve the side lobes except some 
hint of the first side lobe in both beams.}

\chihway{We note that the main beam size from the sun measurement systematically smaller than that from the drone measurement. 
A few factors could contribute to this discrepancy: First, the telescope had an elevation angle of 90$^{\circ}$ in the 
drone measurement and 19$^{\circ}$ in the sun measurement. This means that the mechanical structure could differ due 
to gravity and the level of ground pickup will be larger in the drone measurements. In addition, the drone signal was 10 dB 
larger than the sun signal, which suggested that any non-linear response from the instrument may also cause the two 
beams to produce different measurements. Quantifying exactly how much these contributes to the difference would however 
require more data taken over a longer period of time.}

\subsubsection{Comparison of different measurement methods}
\label{sec:procon}

We have shown above that the two measurements of the beam are consistent, confirming that there are no unknown 
conceptual issues in using the drone to calibrate radio telescope beams. 
However, there are some fundamental differences between the two different approaches of beam calibration. We 
summarise in \Tref{tab:procon} the pros and cons of the two methods described above, together with other known 
techniques not covered in this paper. 
Overall, the drone measurement provides a more controllable way to calibrate the beam, and has potential to 
perform a broader range of calibrations (radiometric calibration, beam as a function of elevation/azimuth, polarisation 
measurements etc.). On the other hand, the drone measurement becomes more challenging going to larger telescopes 
where the far-field is much further away from the antenna. Possible workarounds of this issue include measuring the beam with the 
telescope pointing at a smaller elevation angle, or simply probing the near-field beam instead.    

\begin{deluxetable*}{lllll}
\tablewidth{0pt}
\centering
\tablecaption{Comparison between different approaches of radio beam calibration.} 
\tablehead{ Characteristic & Drone & Sun & Weaker astronomical sources$^{a}$ & Available satellite$^{b}$} 
\startdata
Controllable flux, position and time & yes & no & no & no \\ 
Radiometric calibration & yes & limited$^{c}$ & yes  & yes \\
Point source          & yes & telescope-dependent$^{d}$ & yes & yes \\
At infinity                & no    & yes          & yes    & yes                    \\
SNR                       & controllable & lower & lowest & high         \\
Wavelength range  & broad & broad & depend on source & limited \\
Free                       & no  & yes & yes & yes 
\enddata
\tablenotetext{a}{For example, moon, Cassiopeia A, Taurus A, Cygnus A, and Virgo A.}
\tablenotetext{b}{Locations of the satellites in the sky need to be known beforehand.}
\tablenotetext{c}{The flux and size of the sun varies with time, thus a good model of the sun is needed for 
accurately accounting its size and for radiometric calibration.}
\tablenotetext{d}{Depending on the resolution of the telescope, the sun can be resolved in some cases.}
\label{tab:procon}
\end{deluxetable*}
\vspace{0.2in}

\section{Conclusions}
\label{sec:conclusion}

In this paper, we describe a novel technique of calibrating the beam of radio telescopes using drones. The advantage of 
this approach is that the calibration is controllable and flexible, which can be customised according to the focus of ones 
science goal. We demonstrate the approach by calibrating a 5m single dish at the Bleien Observatory 
in the 21 cm frequency range. We obtain high quality calibration data that allow us to understand the shape of the beam 
pattern in detail out to the 4th side-lobe. We characterise the wavelength dependency of the beam size, beam efficiency 
and aperture efficiency. The measurements are compared with more conventional measurements using the sun and the 
results are broadly consistent within measurement errors. We discussed the challenges in this experiment. Future 
improvements to the current experiment include: 
\begin{itemize}
\item \textbf{Drone flight pattern design:} The equal-spaced-grid flight pattern used in this work is not optimal, as there 
is more information in the centre of the beam. Using for example, an adaptive grid with varying spacing would be
more effective in mapping the beam pattern.  
\item \textbf{Drone positioning system:} One main error in our measurement comes from the inaccuracy of the 
GPS positioning system. This can be improved by using publicly available GPS augmentation systems or other 
techniques.
\item \textbf{Characterisation of telescope:} Our understanding of the telescope geometry in this work is based 
on old drawings that could be outdated. Measuring more precisely the geometry of the telescope's mechanical 
structure is important for better modelling. 
\item \textbf{Characterisation of horn feed:} Measuring the beam of the horn alone in the lab would help 
disentangle the beam of the telescope from the total beam measured. This would also help us understand how 
the horn is illuminating the dish, and would be essential for further radiometric calibrations.
\item \textbf{Modelling:} \chihway{In \Aref{sec:grasp} we use simple antenna modelling to help understand qualitative features 
in our measurements. However, in order to make quantitative comparisons between measurement and model, 
we require a more advanced software package.}
\end{itemize}  

One of the science drivers for developing new, controllable beam calibration techniques for single-dish 
telescopes comes from the stringent requirement on the knowledge of the telescope beam for cosmological 
HI intensity mapping.
Small uncertainties in the beam would introduce undesirable systematics in the cosmological measurements. 
This work provides a practical solution to the challenge by building a controllable artificial calibration source. This is 
achieved by combining commercial drone technology, well designed experiment setup and careful post processing 
of the data. 

\section*{Acknowledgement} 

We thank Sebastian Seehars, Joel Akeret, Aseem Paranjape, Vinzenz Vogel, Oliver Bichsel, Armin Grün and 
Arnold Benz for useful discussions during the experiment. We thank Mr. Christian Schmid and Mr. Bruno Flück from 
the Koptershop for 
technical assistance in designing and flying the drone. We thank Alexander Zvyagin for designing the noise transmitter 
mechanics and harnessing. We thank the mechanical workshop at ETH for manufacturing all required mechanical parts 
for the noise transmitter and Franz Kronauev for fabricating the antenna. We also thank OFCOM for providing the 
transmission license (\# 1000360868) for the drone flights. 


\appendix

\section{Antenna modelling}
\label{sec:grasp}

\setcounter{table}{0}
\numberwithin{table}{section}

\setcounter{figure}{0}
\numberwithin{figure}{section}

In this appendix we use the software package \texttt{GRASP}\footnote{\url{http://www.ticra.com/products/software/grasp}} 
to produce a simple model of the expected beam pattern. \texttt{GRASP} is a standard engineering software used in 
designing reflector antennas. The main tuneable parameters we use in the software are the EM wave frequency, 
reflector geometry (surface type, diameter, f/D, offset) and horn feed geometry (taper angle, taper, polarisation). We 
are not able to add mechanical structures such as the supporting struts.

The geometric information of the Bleien 5m dish can be obtained from the drawing of the telescope from the time of 
construction in the 1970's, but this information does not include modifications on the telescope that have been done over 
the years. Uncertainty in the horn feed geometry is also present, as the current setup consists only of a low-cost cylindrical 
horn with a wire receiver mounted within. The horn was not designed to precisely match the beam of the telescope. We 
choose the default instrument parameters for our antenna model to be those 
listed in \Tref{tab:grasp_param}. These are set according to the telescope drawing described above and an approximate 
model of the cylindrical horn described in \citet{1984}. In our case, the dish is over-illuminated by the horn. 
 
\begin{figure}
  \begin{center}
  \includegraphics[scale=0.6]{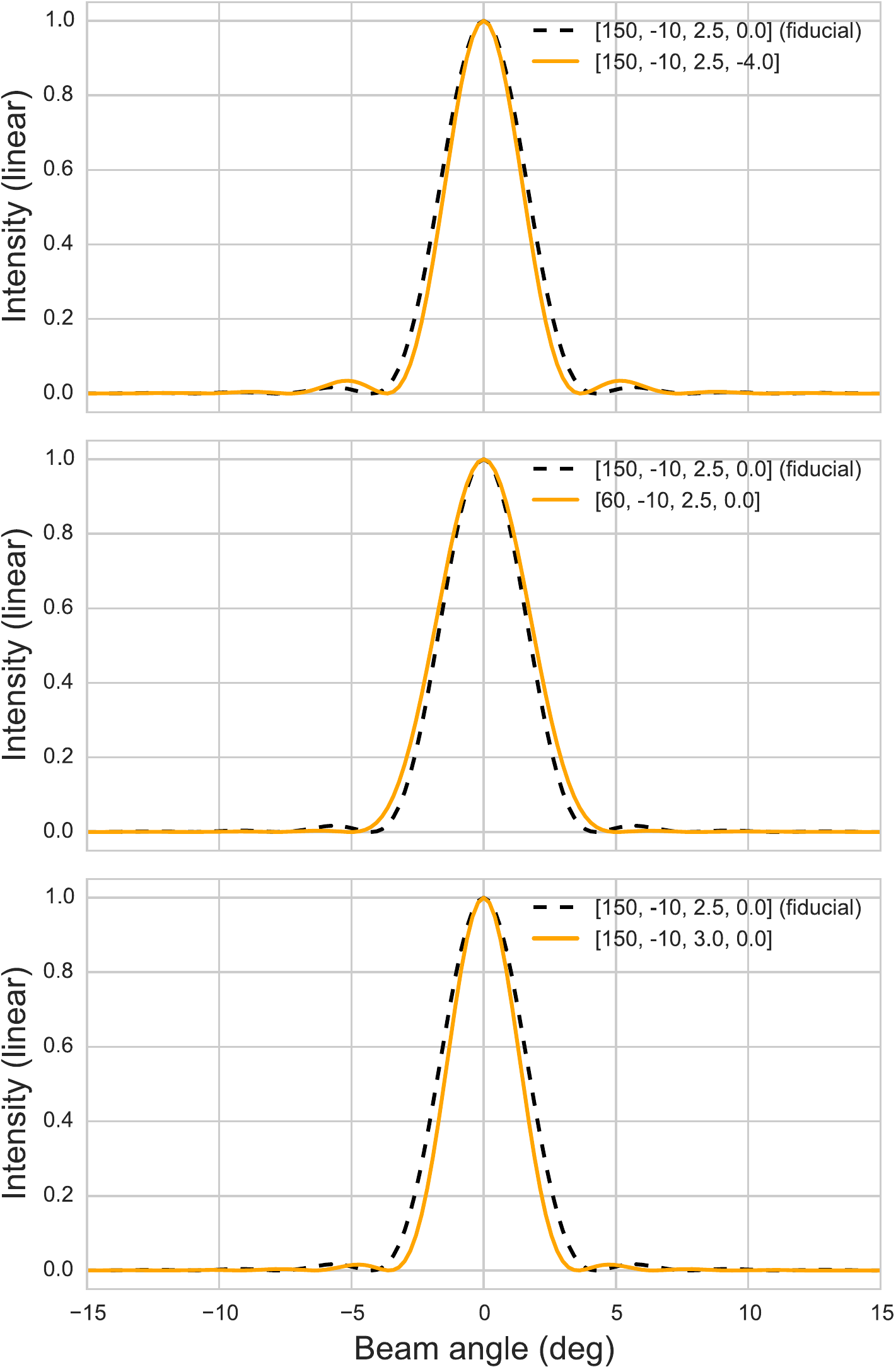} 
  \end{center}
  \caption{\chihway{Beam patterns from \texttt{GRASP} when various instrument parameters are varied. 
  The black dashed curve in all panels are the same, and show the fiducial beam pattern according 
  to \Tref{tab:grasp_param}. The three rows show the effect of varying defocus, tapering angle and the dish size, 
  respectively. The four numbers listed in the legend are [taper angle, taper, radius (m), defocus (cm)] and describe the 
  change in parameters.}}
\label{fig:1d_grasp}
\end{figure}

\chihway{We use \texttt{GRASP} mainly to understand the effect on the beam pattern when a more realistic bean of the 
feed horn is included. We also explore the impact of changing different parameters. 
\Fref{fig:1d_grasp}, for example, shows the 
effect of the beam shape when we perturb the tapering angle, the defocusing and the dish diameter from the fiducial 
setting in \Tref{tab:grasp_param}. We find that change in these parameters have an impact on the positions of the 
peaks/nulls as well as the relative height of the different peaks. 
These qualitative changes are consistent with that described in \citet{2007ASSL..348.....B}. A more sophisticated 
model is needed to address more subtle changes in the shape of the beam.}

\begin{deluxetable}{ll}
\tablewidth{0pt}
\centering
\tablecaption{Default parameters for \texttt{GRASP} antenna modelling.} 
\tablehead{Parameter & Value} 
\startdata
Reflector type & parabolic \\
Reflector f/D & 0.507                 \\
Reflector diameter   & 5m    \\
Frequency & 1114 MHz \\
Feed taper angle  & 150$^{\circ}$ \\
Feed taper  & -10 dB   \\
Defocus & 0
\enddata
\label{tab:grasp_param}
\end{deluxetable}

\end{document}